\documentclass[aps,twocolumn,groupedaddress,floatfix]{revtex4-1}
\usepackage{amssymb,amsmath}
\usepackage{graphicx}
\usepackage{subfigure}
\usepackage[english]{babel}
\usepackage{float}
\usepackage{color}

\usepackage{lipsum}
\begin{document}
\newcommand{\be}{\begin{equation}}
\newcommand{\ee}{\end{equation}}
\newcommand{\rojo}[1]{\textcolor{red}{#1}}

\title{The Fractional Discrete Nonlinear Schr\"{o}dinger Equation}

\author{Mario I. Molina}
\affiliation{Departamento de F\'{\i}sica, Facultad de Ciencias, Universidad de Chile, Casilla 653, Santiago, Chile}

\date{\today }

\begin{abstract} 
We examine a fractional version of the discrete Nonlinear Schr\"{o}dinger (dnls) equation, where the usual discrete laplacian is replaced by a fractional discrete laplacian. This leads to the replacement of the usual nearest-neighbor interaction to a long-range intersite coupling that decreases 
asymptotically as a power-law. For the linear case, we compute both, the spectrum of plane waves and the mean square displacement of an initially localized excitation in closed form, in terms of regularized hypergeometric functions, as a function of the fractional exponent. In the nonlinear case, we compute numerically the low-lying nonlinear modes of the system and their stability, as a function of the fractional exponent of the discrete laplacian. The selftrapping transition threshold of an initially localized excitation shifts to lower values as the exponent is decreased and, for a fixed exponent and zero nonlinearity, the trapped fraction remains greater than zero.

\end{abstract}

\maketitle

{\em Introduction}. Let us consider the discrete nonlinear Schr\"{o}dinger (dnls) equation that describes the motion of a nonlinear excitation propagating along a discrete lattice\cite{kevrekidis,dst,eilbeck}
\be
i {d C_{n}\over{d t}} + V (C_{n+1}+C_{n-1}) + \chi |C_{n}|^2 C_{n}=0\label{eq:1}
\ee
This equation has been used to describe the propagation of excitations in a deformable medium\cite{molecule1,molecule2}, transversal propagation of light in waveguide arrays\cite{optics1,optics2,optics3,optics4}, dynamics of Bose-Einstein condensates inside coupled magneto-optical traps\cite{BE1,BE2}, self-focusing and collapse of Langmuir waves in plasma physics\cite{plasmas1,plasmas2} and description of rogue waves in the ocean\cite{rogue}, among others. 
Its most distinctive feature is the existence of localized solutions, termed discrete solitons, with families of stable and unstable modes that, in general, exist above a certain nonlinearity strength. The dynamics of the dnls equation shows the existence of a selftrapping transition\cite{MM_GTS,MM_GTS2} for an initially localized excitation, as well as a degree of mobility (in 1D) across the lattice\cite{optics3}. Because of all these properties, the dnls equation has  gained the status of a paradigmatic equation that describes the propagation of excitations in a nonlinear medium in a variety of different physical scenarios. 

On the other hand, the topic of fractional derivatives has gained increased attention in the last years. It started with the observation that a usual integer-order derivative could be extended to a fractional-order derivative, $(d^n/dx^n)\rightarrow (d^{s}/d x^s)$, for real $s$. For instance, the $s$-th derivative of a function $f(x)$ can be formally expressed as\cite{fractional1,bjwest,fractional2}
\be
\left({d^{s}\over{d x^{s}}}\right) f(x) = {1\over{\Gamma(1-s)}} {d\over{d x}} \int_{0}^{x} {f(x')\over{(x-x')^{s}}} dx'\label{eq:2}
\ee
\vspace{0.25cm}
for $0<s<1$. For the case of the laplacian operator $\Delta=\partial^2/\partial {\bf r}^2$, its fractional form $(-\Delta)^s$ can be expressed (in one dimension) as\cite{continuous laplacian}
\be
(-\Delta)^s U(x) = C_{s} \int {U(x)-U(y)\over{|x-y|^{n+2 s}}} dy\label{eq:3}
\ee
where,
\[
C_{s}={4^s \Gamma(s+(1/2))\over{\sqrt{\pi}|\Gamma(-s)|}},\label{eq:4}
\]
$\Gamma(x)$ is the gamma function and $0<s<1$ is called the order of the laplacian. This form of the fractional laplacian have proved useful in applications to fluid mechanics\cite{30,35}, fractional kinetics and anomalous diffusion\cite{71,86,101}, strange kinetics\cite{82}, fractional quantum mechanics\cite{64,65}, Levy processes in quantum mechanics\cite{75}, plasmas\cite{2}, electrical propagation in cardiac tissue\cite{20} and biological invasions\cite{9}.

In this work we aim at examining the consequences of the use of a fractional discrete laplacian
on the existence and stability of nonlinear modes of the discrete nonlinear Schr\"{o}dinger (dnls) equation, as well as in the transport of excitations in this system. As we will see, the usual phenomenology of the dnls equation is more or less preserved, although there are changes in the spectrum of plane waves, becoming completely flat (i.e., degenerate) at $s\rightarrow 0$. This flattening tendency also affects the capacity of the system to selftrap nonlinear excitations.
 
{\em The model}. The kinetic energy term in Eq.(\ref{eq:1})$, V (C_{n+1}+C_{n-1})$, is nothing else but a  discretized form of the laplacian $\Delta_{n} C_{n} = C_{n+1}-2 C_{n}+C_{n-1}$, so that 
Eq.(\ref{eq:1}) can be cast as
\be
i {d C_{n}\over{d t}} + 2 V C_{n} + V \Delta_{n} C_{n} + \chi |C_{n}|^2 C_{n}=0.\label{eq:5}
\ee
We wonder now about the effect of replacing $\Delta_{n}$ by its fractional form  $(\Delta_{n})^s$ in Eq.(\ref{eq:5}). The form of this fractional discrete laplacian is given by\cite{discrete laplacian}:
\be
(-\Delta_{n})^s C_{n}=\sum_{m\neq n} K^s(n-m) (C_{n}-C_{m}),\hspace{0.5cm}0<s<1 \label{delta}
\ee
where,
\be
K(m) = C_{s}\  {\Gamma(|m|-s)\over{\Gamma(|m|+1+s)}}.\label{K}
\ee
After replacing Eqs.(\ref{delta}) and (\ref{K}) into Eq.(\ref{eq:5}) and searching for a stationary-state mode $C_{n}(t) = \exp(i \lambda) \phi_{n}$, we obtain the following system of nonlinear difference equations for $\phi_{n}$:
\be
(-\lambda + 2 V )\ \phi_{n} + V \sum_{m\neq n} K^s(n-m) (\phi_{m}-\phi_{n}) + \chi\ \phi_{n}^3=0\label{eq:7}
\ee
where, without loss of generality $\phi_{n}$ can be chosen as real. We see that the immediate effect  of the fractional discrete laplacian is to introduce nonlocal interactions via a symmetric kernel $K^s(n-m)$.

{\em Plane waves}. Let us start by taking $\chi=0$ and looking for solutions of the form $\phi_{n} = A\ \exp(i k n)$. After some simple algebra, one obtains the dispersion relation for plane waves
\be
\lambda(k) = 2V - 4 V \sum_{m=1}^{\infty} K(m) \sin((1/2) m k)^2\label{dispersion}
\ee
or, in closed form,
\begin{widetext}
\be
\lambda(k)=2-{16\ \Gamma(s+(1/2))\over{\sqrt{\pi}\ \Gamma(1+s)}}\Big( 1-\exp(-i k)\ s\ \Gamma(1+s)[\ R(1,1-s,2+s;\exp(-i k))+\exp(2 i k) \ R(1,1-s,2+s;\exp(i k))\ ] \Big)
\ee
\end{widetext}
where $R(a,b,c;z)={}_2 F _{1}(a,b,c;z)/\Gamma(c)$ is the regularized hypergeometric function.

\begin{figure}[t]
 \includegraphics[scale=0.2]{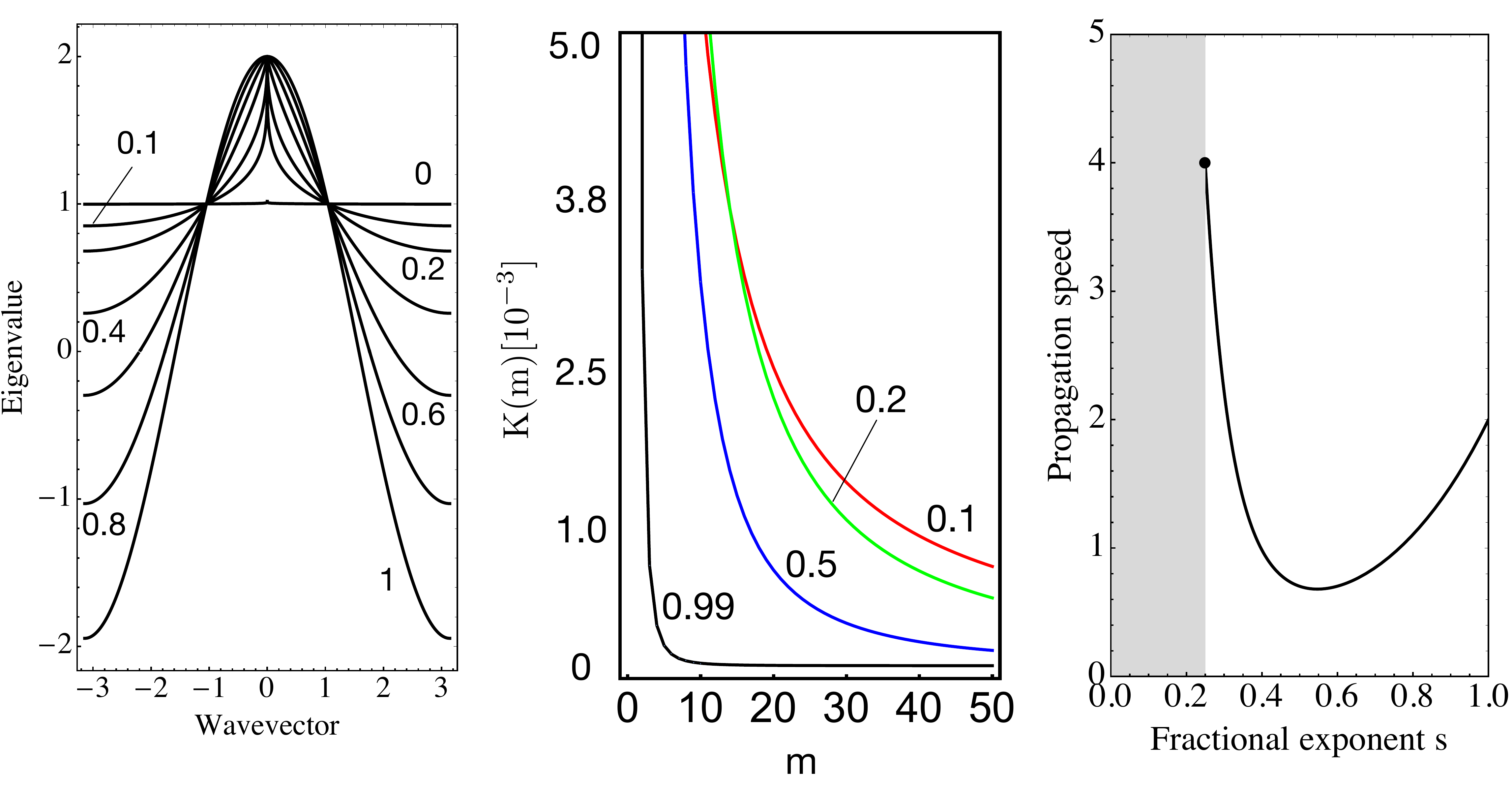}
  \caption{Left: Dispersion relation $\lambda(k)$ for several different fractional exponents $s$ marked on each curve ($V=1$). 
 Middle: Decrease of kernel $K(m)$ with distance for several fractional exponents. 
 Right: propagation speed of an excitation as a function of the fractional exponent. The shaded area denotes values of $s$ where the RMS is not well defined.}
  \label{fig1}
\end{figure}

\begin{figure}[t]
 \includegraphics[scale=0.5]{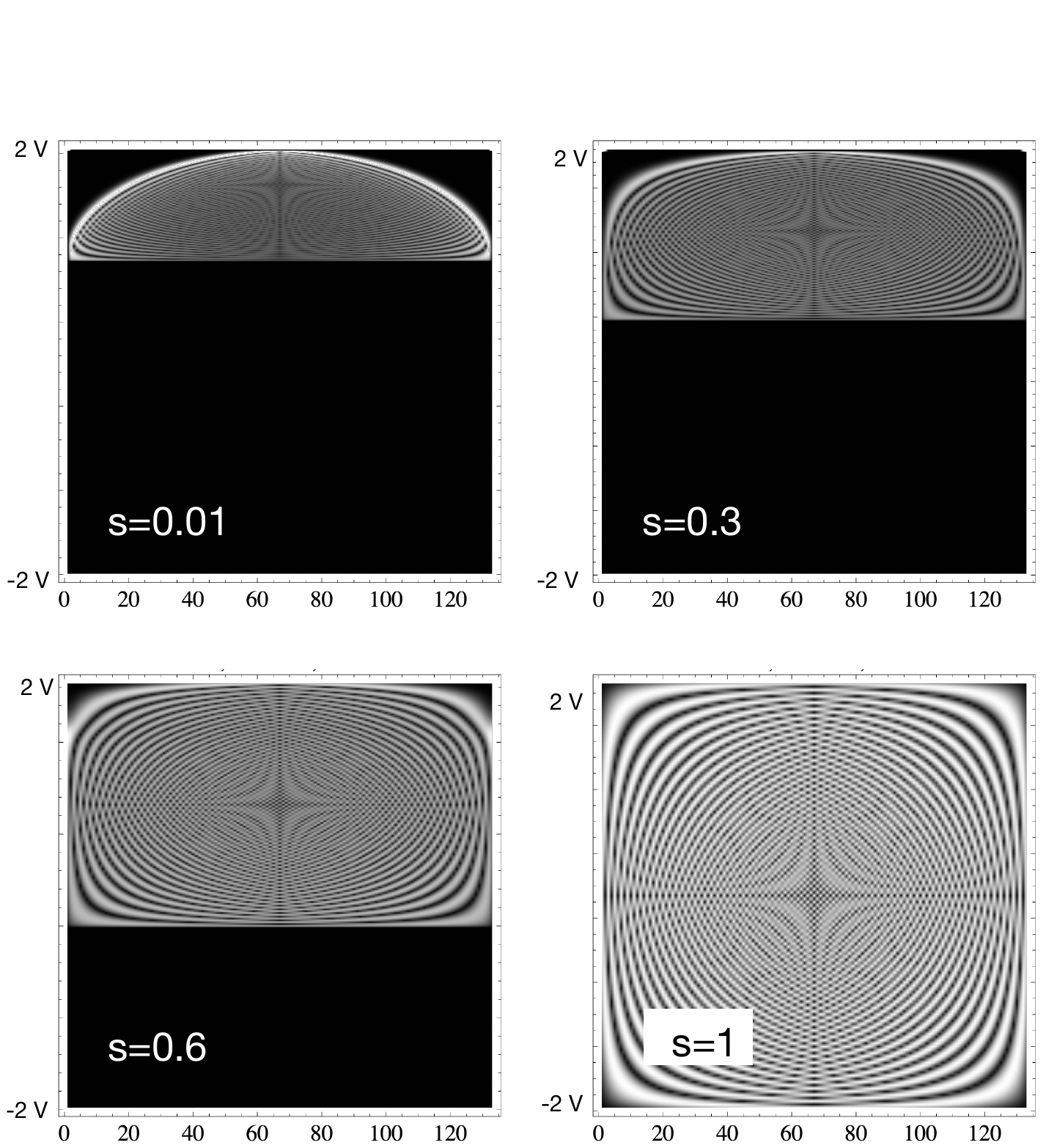}
  \caption{Density plot of the spatial profiles $|\phi_{n}|^2$ of the linear modes ordered according to their eigenvalue. For $s\sim 0\ (s\sim 1)$ the bandwidth is $V\ (4 V)$\ ($N=133$).
  }
  \label{fig2}
\end{figure}

Using $\Gamma(n+s)=\Gamma(n) n^s$, one obtains the asymptotic form $K(m)\rightarrow 1/|m|^{1+ 2 s}$, evidencing a power-law decrease of the coupling with distance. The dispersion $\lambda(k)$ is well-defined for $s>0$. For $s$ approaching unity, we have $\lim_{s\rightarrow 1^{-}} K(m) = \delta_{n,1}$, while near $s=0$ we have $\lim_{s\rightarrow 0^{+}} K(n)\rightarrow \mbox{sign}(n) s/n$. Thus, near $s=1$ the coupling is mainly between nearest neighbors, while near zero it becomes long-ranged. 
Figure \ref{fig1} (left panel)) shows dispersion curves for different fractional exponents $s$, ranging from $s=0^{+}$ up to $s=1^{-}$. The bandwitdth $\Delta\lambda=\lambda(0)-\lambda(\pm \pi)$ changes with $s$, increasing from a minimum value of $V$ (at $s=0$) up to $4 V$ (at $s=1$). The decrease of the kernel $K(m)$ with distance is also shown in Fig.\ref{fig1} (middle panel). We note that as $s$ decreases, the range of $K(m)$ increases causing an increase in the effective range of the coupling among sites. In the limit $s\rightarrow 0$, all sites become similarly  coupled, and the resulting system is similar to what is known in the literature as a {\em simplex}\cite{simplex1,simplex2}. We will come back to this point later on.
Figure \ref{fig2} shows density plots for the mode profiles, for several values of the fractional exponent $s$. Here, for a given $s$, we stack the mode profiles one after the other according to their eigenvalues. As we can see, and in consonance with Fig.\ref{fig1}, the bandwidth $\Delta\lambda$ increases from a minimum value of $V$ (at $s=0$) to a value of $4 V$ (at $s=1$). 

{\em Root mean square (RMS) displacement}. A way to quantify the transport across the system is by means of the root mean square (RMS) displacement of an initially localized excitation. For a periodic lattice with a well-defined dispersion relation that satisfies the general conditions 
$\lambda(-k)=\lambda(k)$, $\lambda(k)=\lambda(k+2 \pi q), q\in Z$, and $(\partial/\partial k)\lambda(k)|_{k=0}=(\partial/\partial k)\lambda(k)|_{k=\pm \pi}=0$, it can be proven that the mean square displacement
\be
\langle n^2\rangle = \sum_{n} n^2 |\phi_{n}(t)|^2/\sum_{n} |\phi_{n}(t)|^2
\ee
is always ballistic and given by \cite{Molina_Martinez}
\be
\langle n^2 \rangle = \left[ {1\over{2 \pi}} \int_{-\pi}^{\pi} \left( {d \lambda(k)\over{d k}}\right)^2\ dk\right]\ (V t)^2,
\ee
for a  localized initial excitation ($\phi_{n}(0)=\delta_{n,0}$).
Using the form of $\lambda(k)$ given by Eq.(\ref{dispersion}), one obtains,
\be
\langle n^2 \rangle = 2 \sum_{m=1}^{\infty} (m K^{s}(m))^2 (V t)^2.
\ee
Using the asymptotic form $K^{s}(n)\rightarrow 1/n^{1+ 2s}$, we have $(n K^{s}(n))^2\rightarrow 1/n^{4 s}$. This implies that $\langle n^2 \rangle$ is well-defined for $s>1/4$ only.
The square of the `speed' $\langle n^2 \rangle/(V t)^2$ can be written in closed form as
\begin{multline*}
{\langle n^2 \rangle\over{(V t)^2}}=\left({1\over{\pi}}\right) 2^{4s-1}\ s\ \Gamma(s+(1/2))^2\\
\times \left\{ {1\over{\Gamma(1+s)^2}} + {8s(s-1)^2\over{\Gamma(3+s)^2}}\ {}_p F _{q} (\{a\},\{b\},1)\right\}\end{multline*}
where, $\{a\}=\{3,2-s,2-s\}, \{b\}=\{3+s,3+s\}$ and 
${}_p F _{q}(\{a\},\{b\};z)$ is the generalized hypergeometric function. Figure \ref{fig1} (right panel) shows this speed versus the fractional exponent $s$ for $0<s<1$. At $s=1/4$ the speed is $4$ while at $s=1$, the speed is $2$. Minimum propagation speed of $0.680364$ is attained at $s=0.5457$.

\begin{figure}[t]
 \includegraphics[scale=0.18]{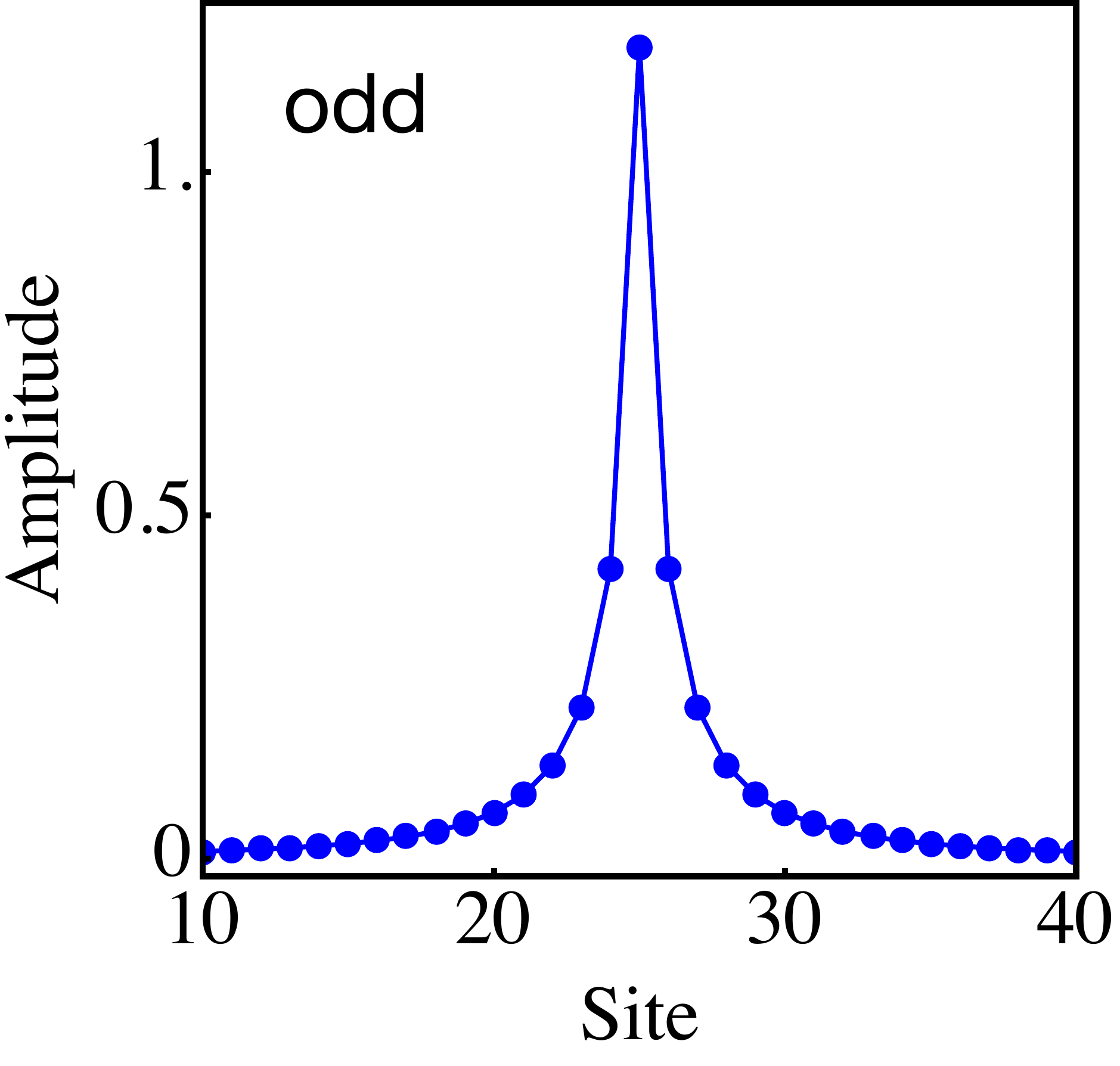}
 \includegraphics[scale=0.18]{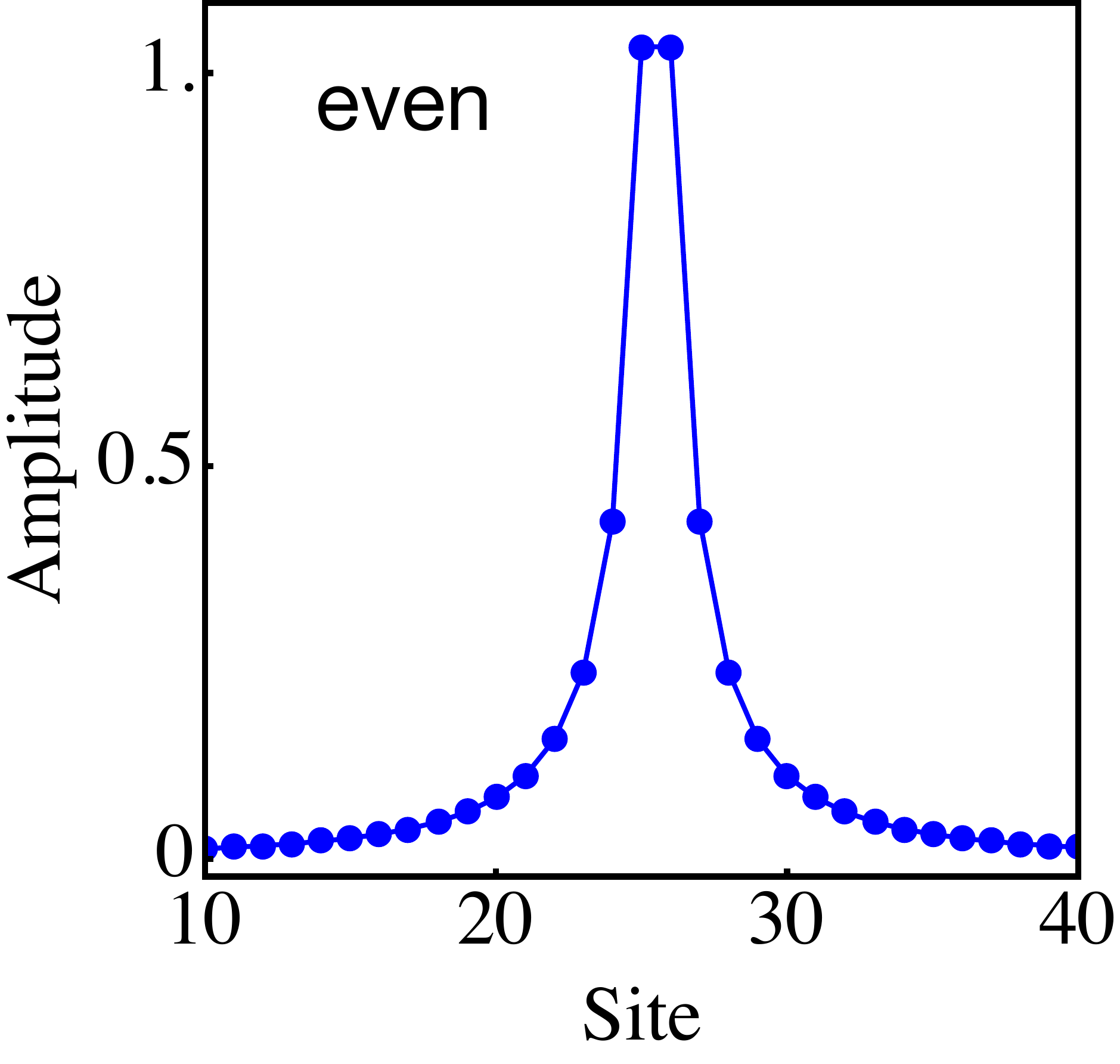}\\
 \includegraphics[scale=0.18]{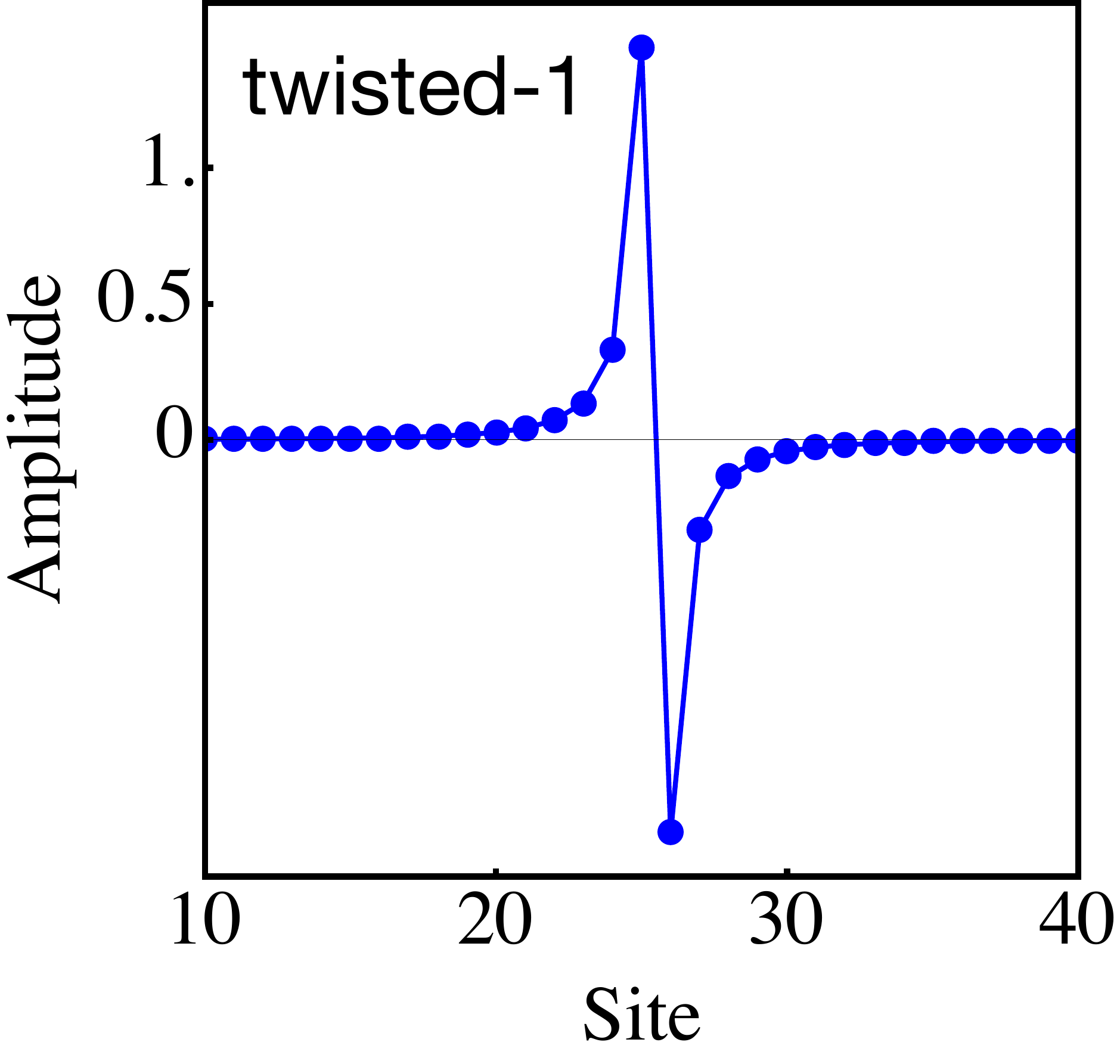}
 \includegraphics[scale=0.18]{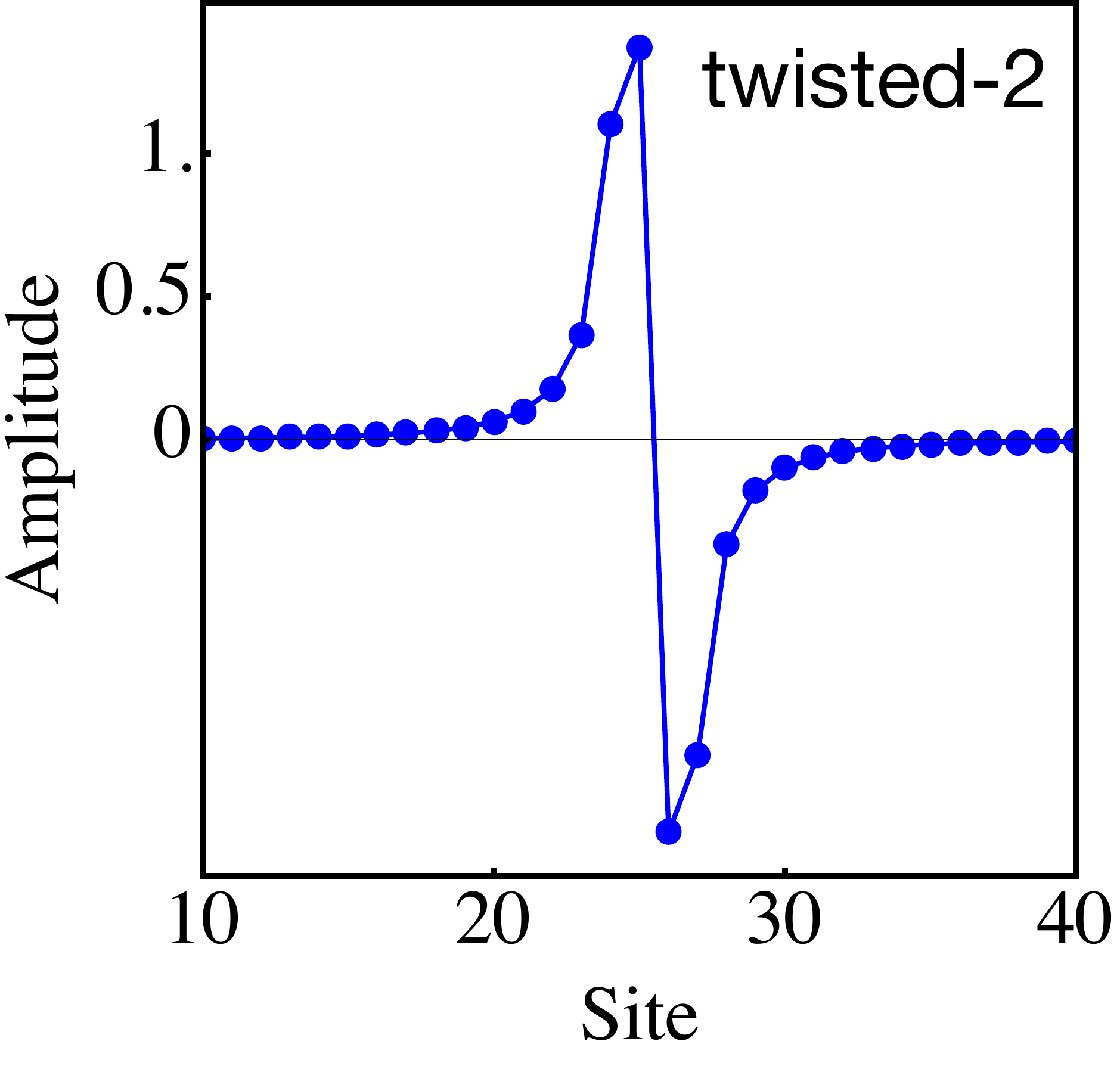}
  \caption{Examples of spatial profiles of some low-lying bulk modes, for $s=1/2$. (a) `odd' mode, (b) `even' mode (c), `twisted-1' mode, and `twisted-2' mode. ($N=51, \lambda=2.6$)
  }
  \label{fig3}
\end{figure}

\noindent
{\em Nonlinear modes}.\ Let us go back to the stationary Eq.(\ref{eq:7}) for $\chi\neq 0$. It constitutes a system of coupled nonlinear equations, with a nonlocal coupling. The form of the nonlinear term is typical for nonlinear optical waveguide arrays and also in electron propagation in a deformable lattice, in the semiclassical approximation. Numerical solutions are obtained by the use of a multidimensional Newton-Raphson scheme, using as a seed the form obtained from the decoupled limit ($V\rightarrow 0$), also known as the anticontinuous limit. The boundary conditions are open, with lattice sites ranging from $n=1$ up to $n=N$. We will examine two mode  families, ``bulk'' modes, located far from the boundaries and ``surface'' modes located near the beginning (or end) of the lattice.
Figure \ref{fig3} shows spatial 
profiles of some nonlinear bulk modes, for $s=1/2$. The shape of these profiles is similar to the ones found for the dnls case ($s=1$), and are known in the literature as `odd', `even',  `twisted$-1$' and `twisted$-2$'. For other values of $s$ we find profiles modes have that are similar (but no identical) to their dnls counterpart.

Figure \ref{fig4} shows some surface modes for $s=1/2$. These modes, in particular, correspond to those modes whose center is located at the surface, one layer below the surface, two layers, etc.
As expected, when the center of the surface mode is pushed farther and farther from the surface, the mode begin resembling a bulk one. Unlike the bulk case, these modes do not have common names associated with them, so it is better to classify them by the form they adopt in the anti-continuous limit (sites completely decoupled). This form is also the form of the seed used for the Newton-Raphson iteration. Thus, the mode at the very surface is denoted by $(1,0,0,0,\cdots)$, and is obtained by iterating from an initial state where only the boundary site is excited. The mode located one layer below would be labelled as $(0,1,0,0,\cdots)$. 
We have used this notation scheme for all states displayed in Fig.\ref{fig4}.

{\em Stability of nonlinear modes}.\ The linear stability analysis of the nonlinear modes is carried out by the well-known standard procedure which we sketch here for completeness:  
We replace $\Delta_{n}$ with $(\Delta_{n})^s)$ in Eq.(\ref{eq:1}) and insert a perturbed solution $C_{n}(t)=(\phi_{n} + \delta_{n}(t)) \exp(i \lambda t)$ 
(with $|\delta_{n}(t)|\ll |\phi_{n}(t)|$), followed by a linearization procedure where we neglect any higher power of $\delta_{n}(t)$, save for the linear one. Next we decompose $\delta_{n}(t)$ into its real and imaginary parts: $\delta_{n}(t) = x_{n}(t) + i y_{n}(t)$. This leads to a set of coupled real equations:
\begin{figure}[b]
 \includegraphics[scale=0.18]{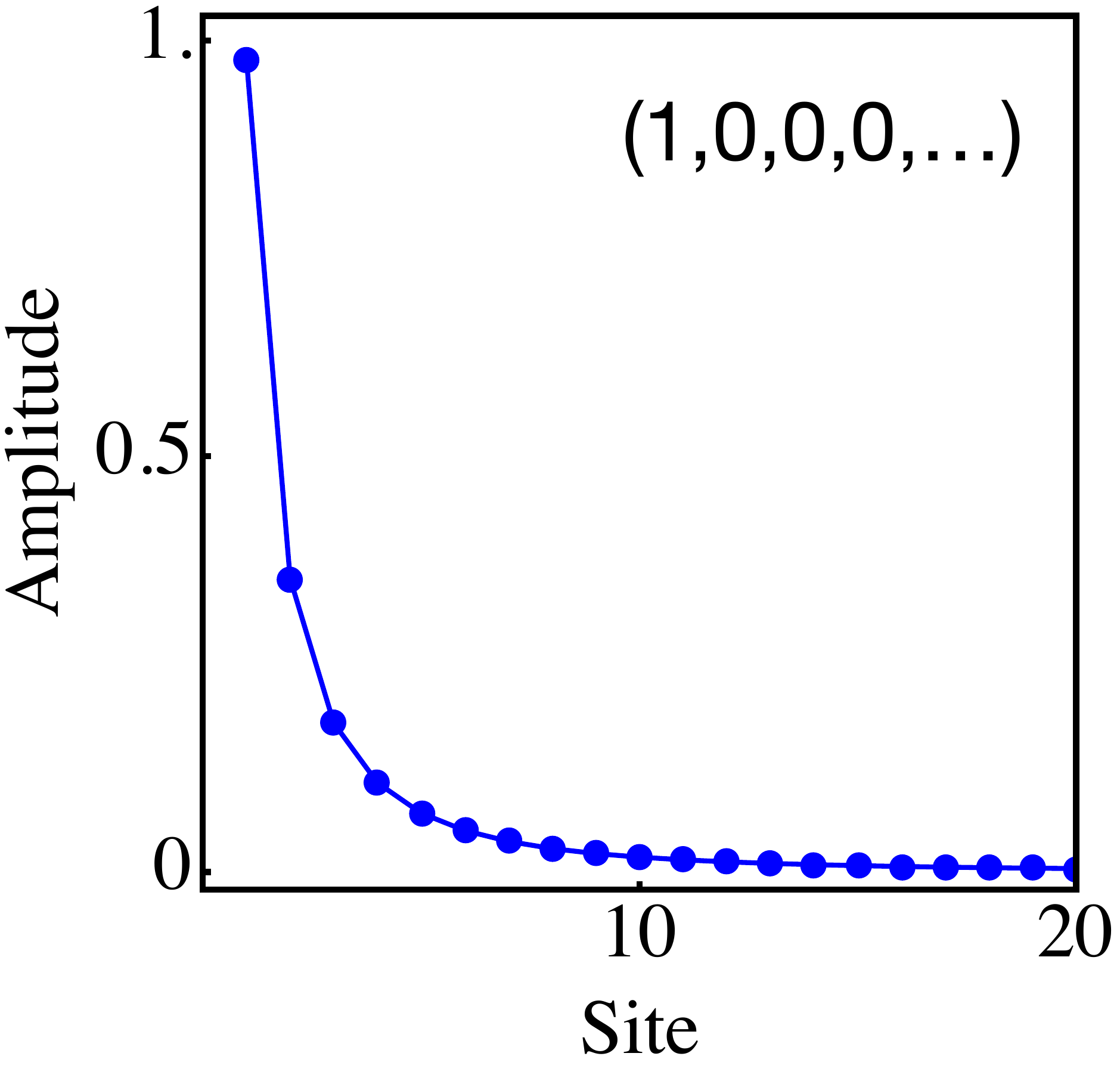}
 \includegraphics[scale=0.18]{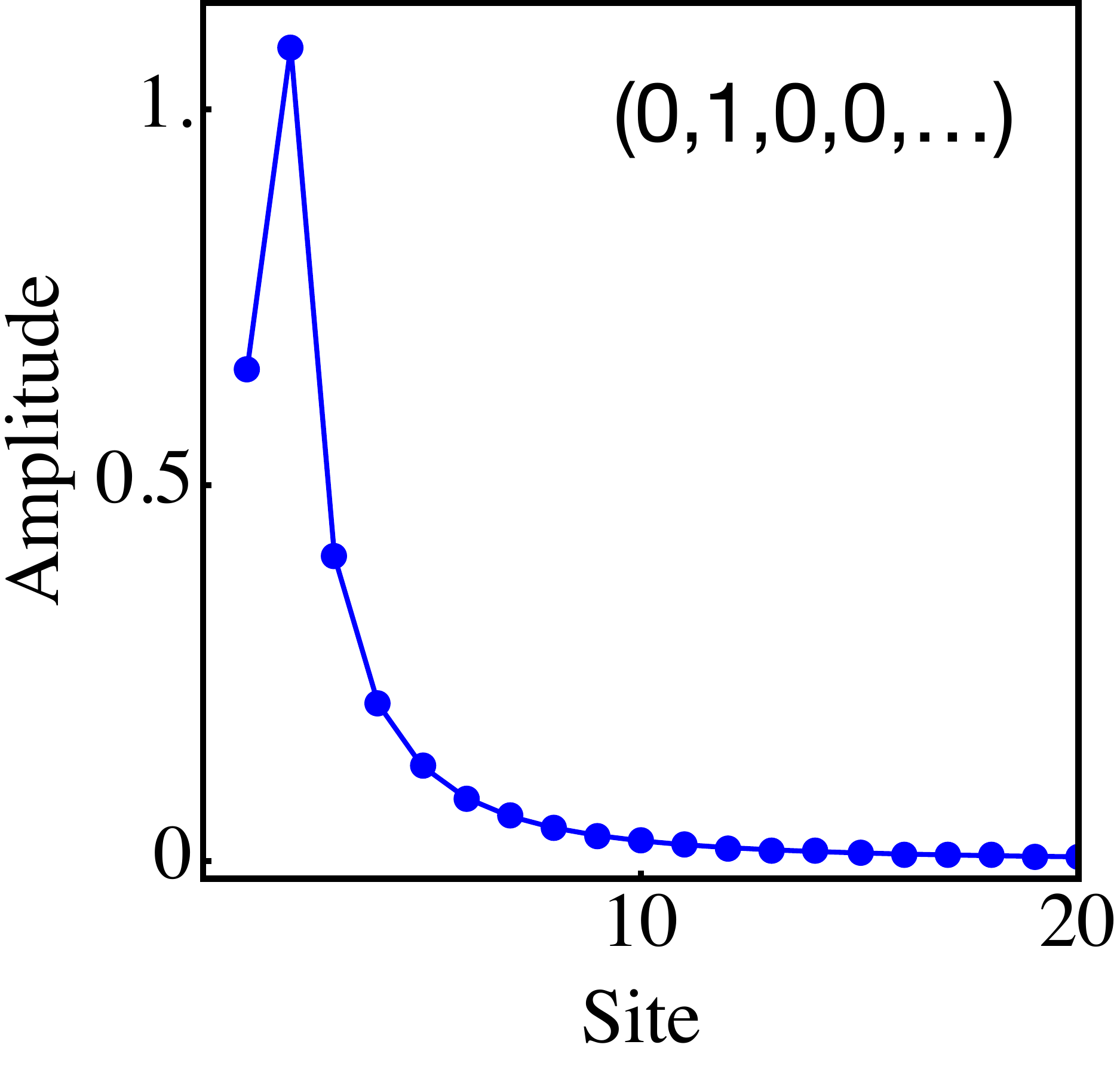}\\
 \includegraphics[scale=0.18]{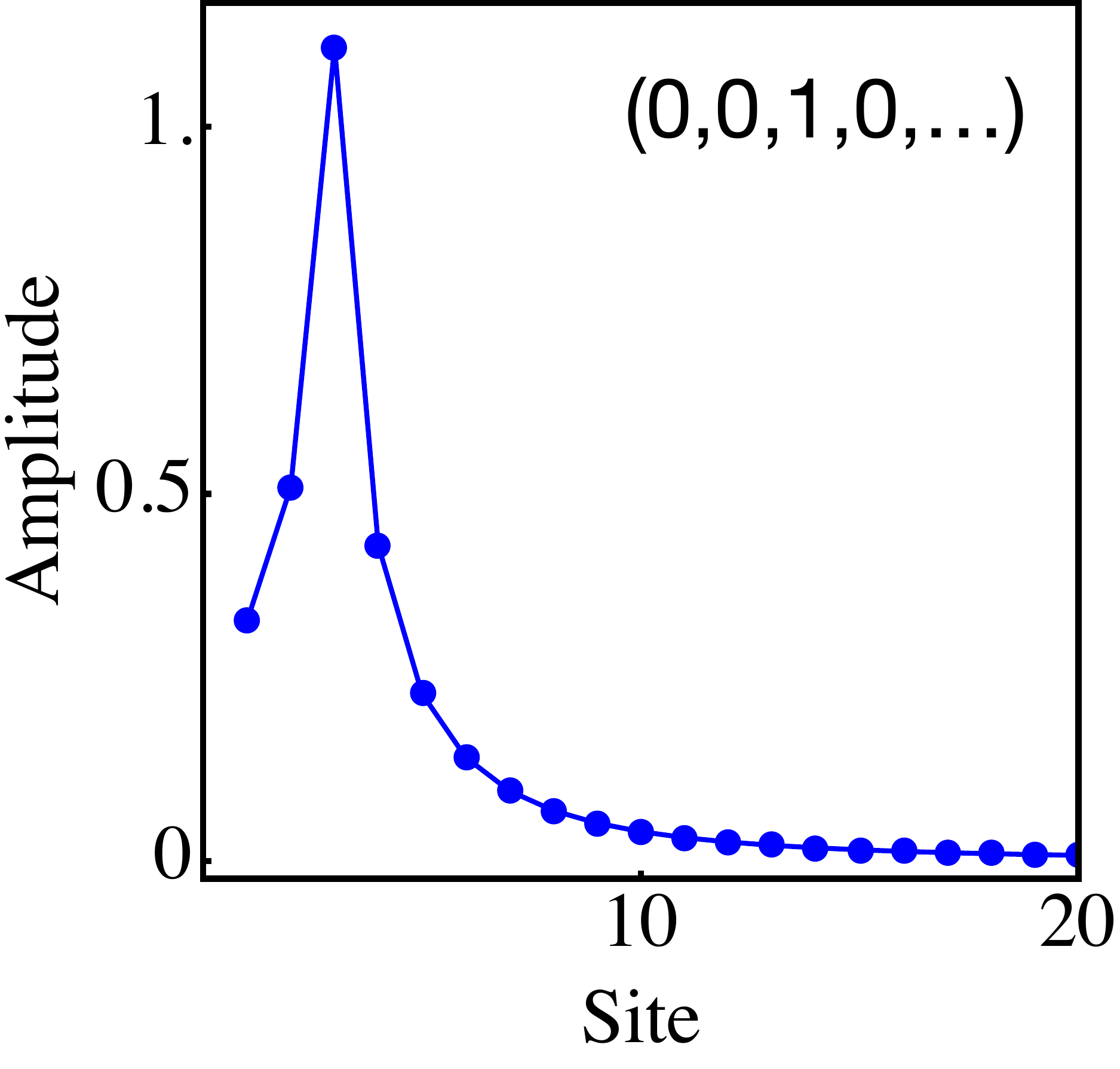}
 \includegraphics[scale=0.18]{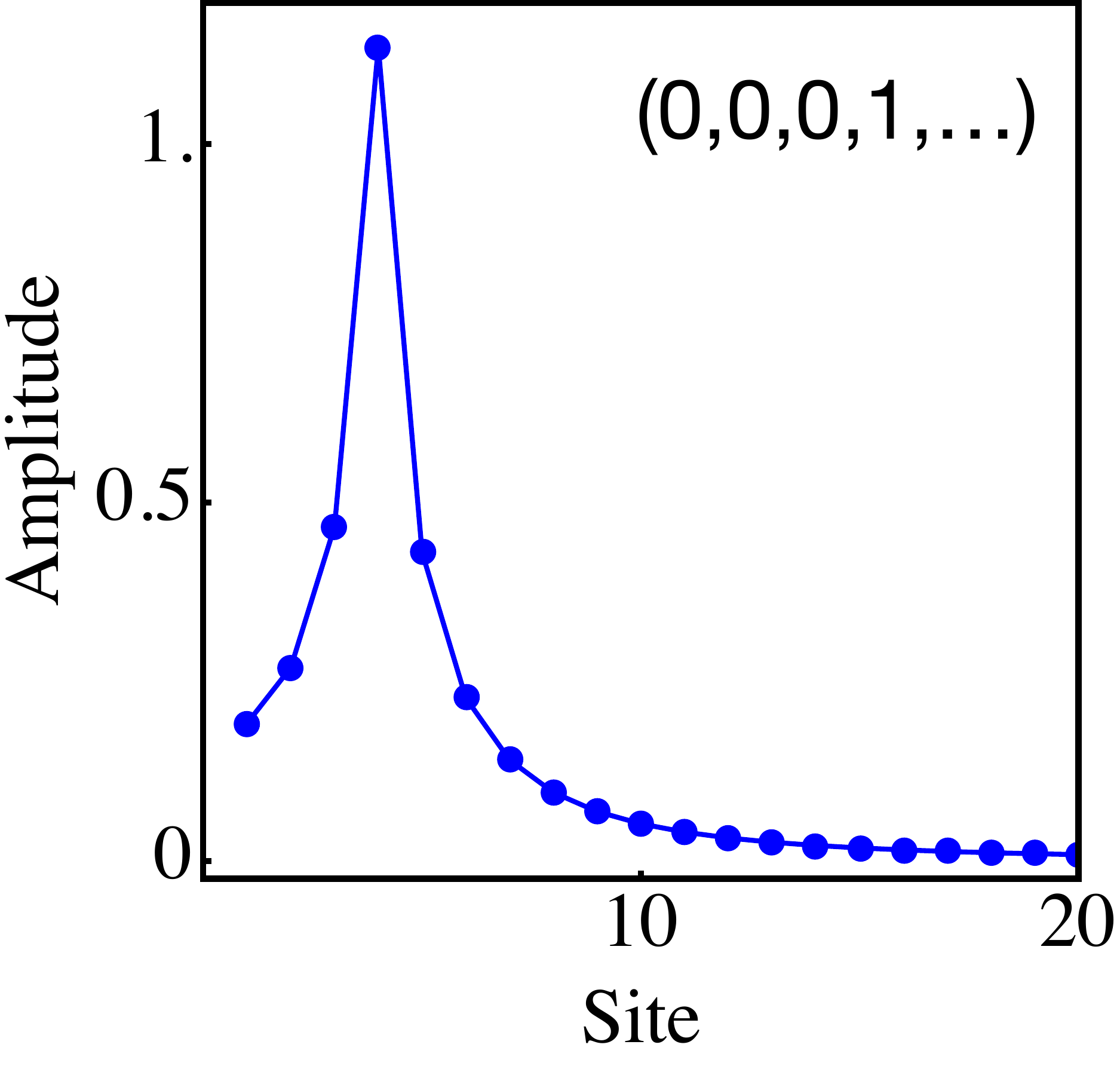}
  \caption{Examples of spatial surface profiles of some low-lying surface modes, for $s=1/2$. Surface is located at $n=1$. The labels on each mode denote the anti-continuous state they originated from. ($N=51, \lambda=2.6$).
  }
  \label{fig4}
\end{figure}
\be
{d^2\over{d t^2}}\ \vec{x} - {\bf A}\ {\bf B}\ \vec{x} = 0, \hspace{0.5cm}{d^2\over{d t^2}}\ \vec{y} - {\bf B}\ {\bf A}\ \vec{y} = 0,\label{stability}
\ee
where $\vec{x}=(x_{1}, x_{2},\cdots x_{N})$ and $\vec{y}=(y_{1}, y_{2},\cdots y_{N})$, and ${\bf A}$ and ${\bf B}$ are matrices given by
\be
{\bf A}_{n m}=K(n-m)+ [-\lambda+2 V+\phi_{n}^2-\sum_{j\neq n} K(n-j)]\ \delta_{n m}
\ee
\be
{\bf B}_{n m}=K(n-m)+ [-\lambda+2 V+3 \phi_{n}^2-\sum_{j\neq n} K(n-j)]\ \delta_{n m}
\ee
From Eq.(\ref{stability}) we see that the linear stability is determined by the eigenvalue spectra of the matrices ${\bf A B}$ and ${\bf B A}$ (both have the same spectrum).
This means determining the largest growth rate, for each mode. This is known as the instability gain, defined as
\be
G = \mbox{Max of}\left\{ {1\over{2}}\left( \mbox{Re(g)}+\sqrt{\mbox{Re(g)}^2 + \mbox{Im(g)}^2}\right) \right\}^{1/2}
\ee
for all $g$, where $g$ is an eigenvalue of ${\bf A B}$ (${\bf B A}$). When $G=0$ the modes is stable; otherwise it is unstable.
\begin{figure}[t]
 \includegraphics[scale=0.3]{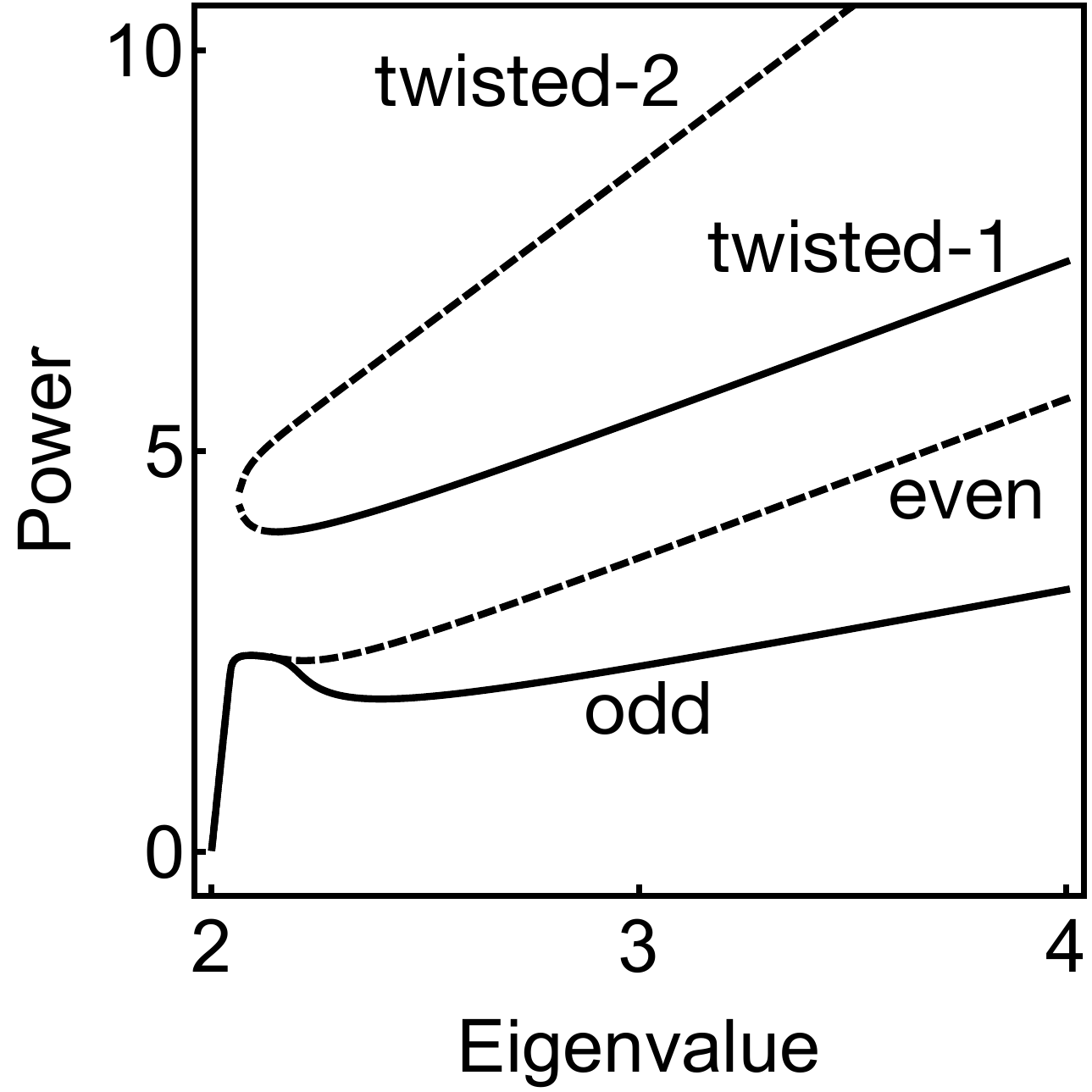}
 \includegraphics[scale=0.3]{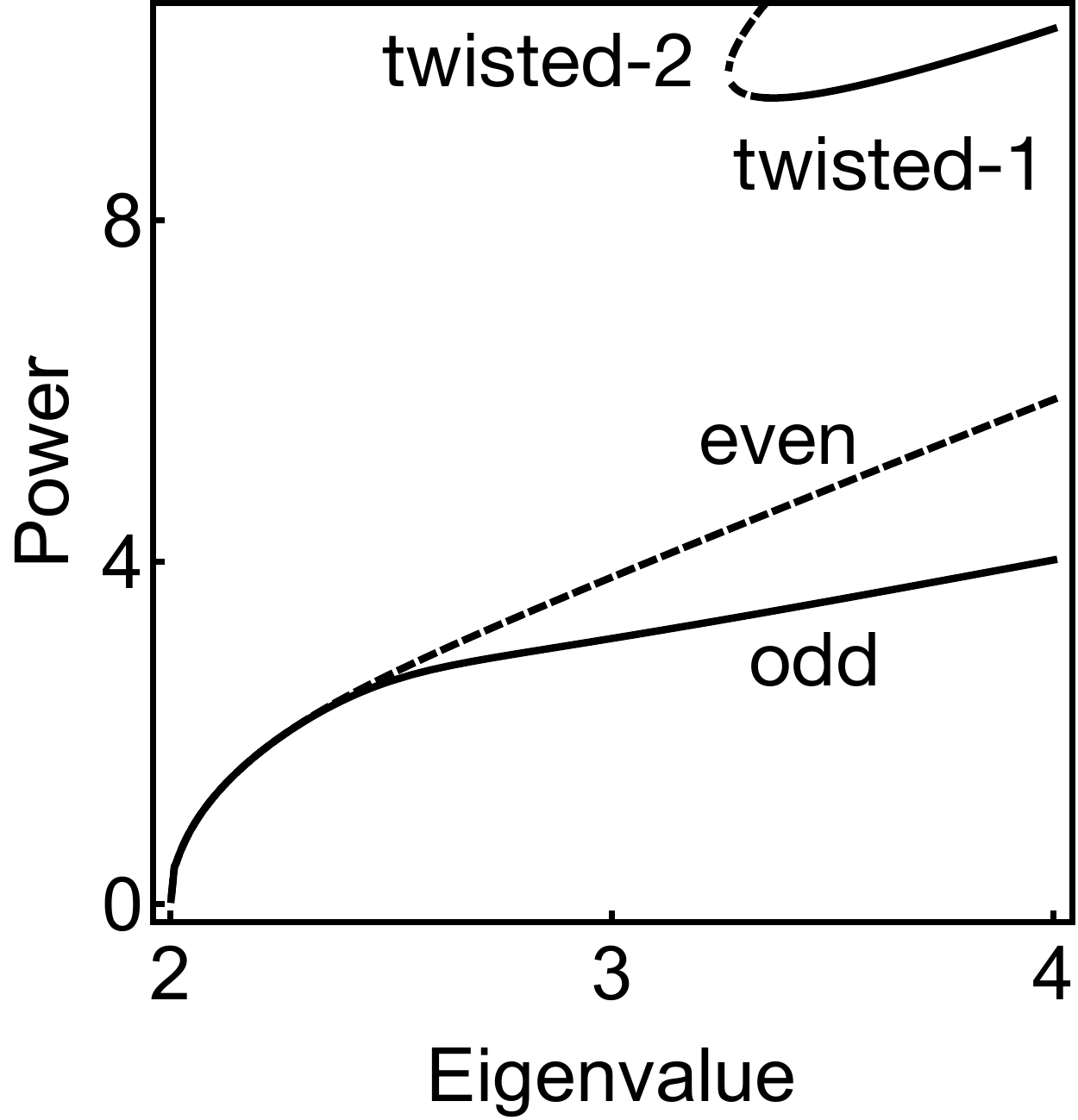}\\
 \includegraphics[scale=0.3]{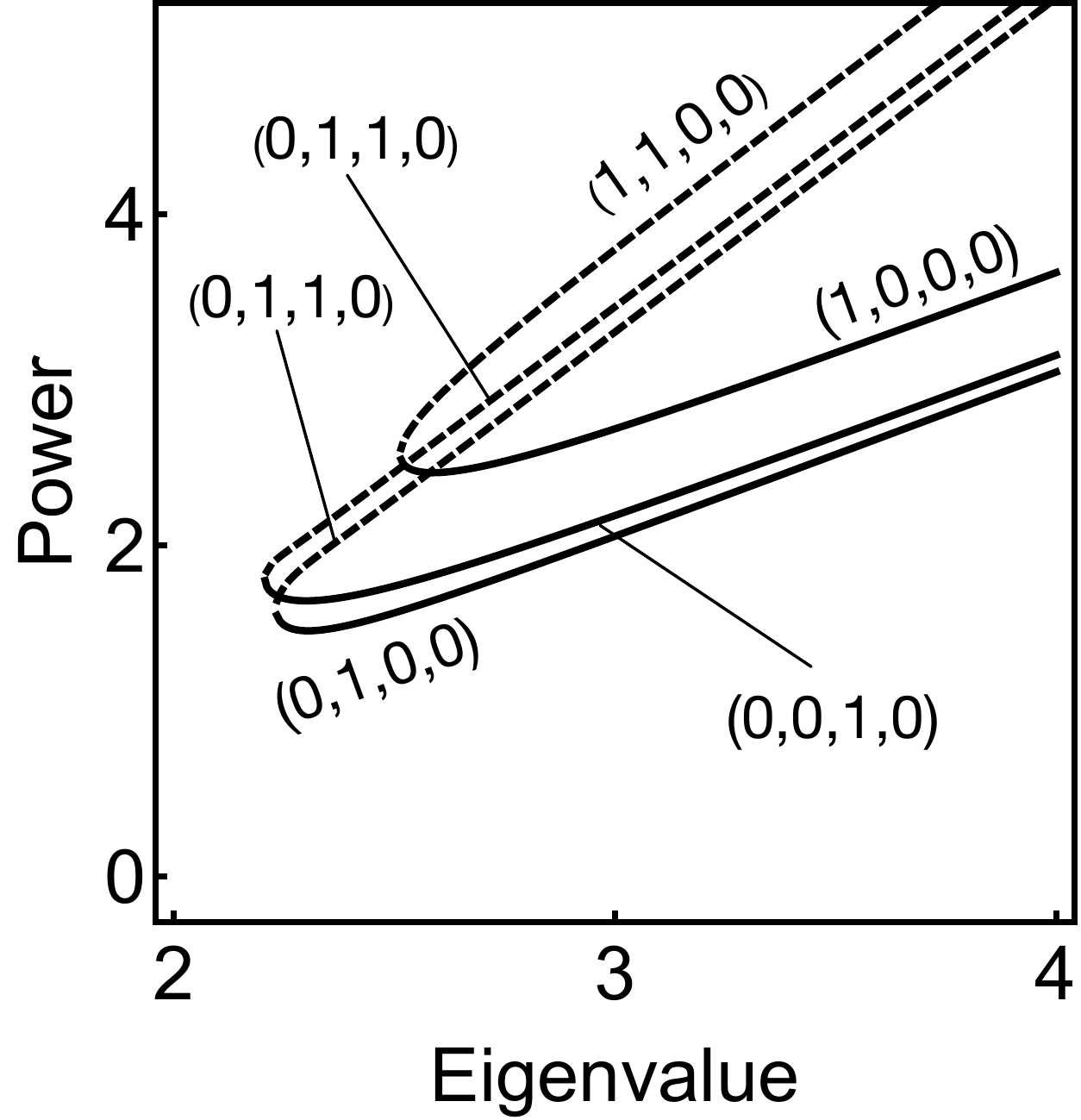}
 \includegraphics[scale=0.3]{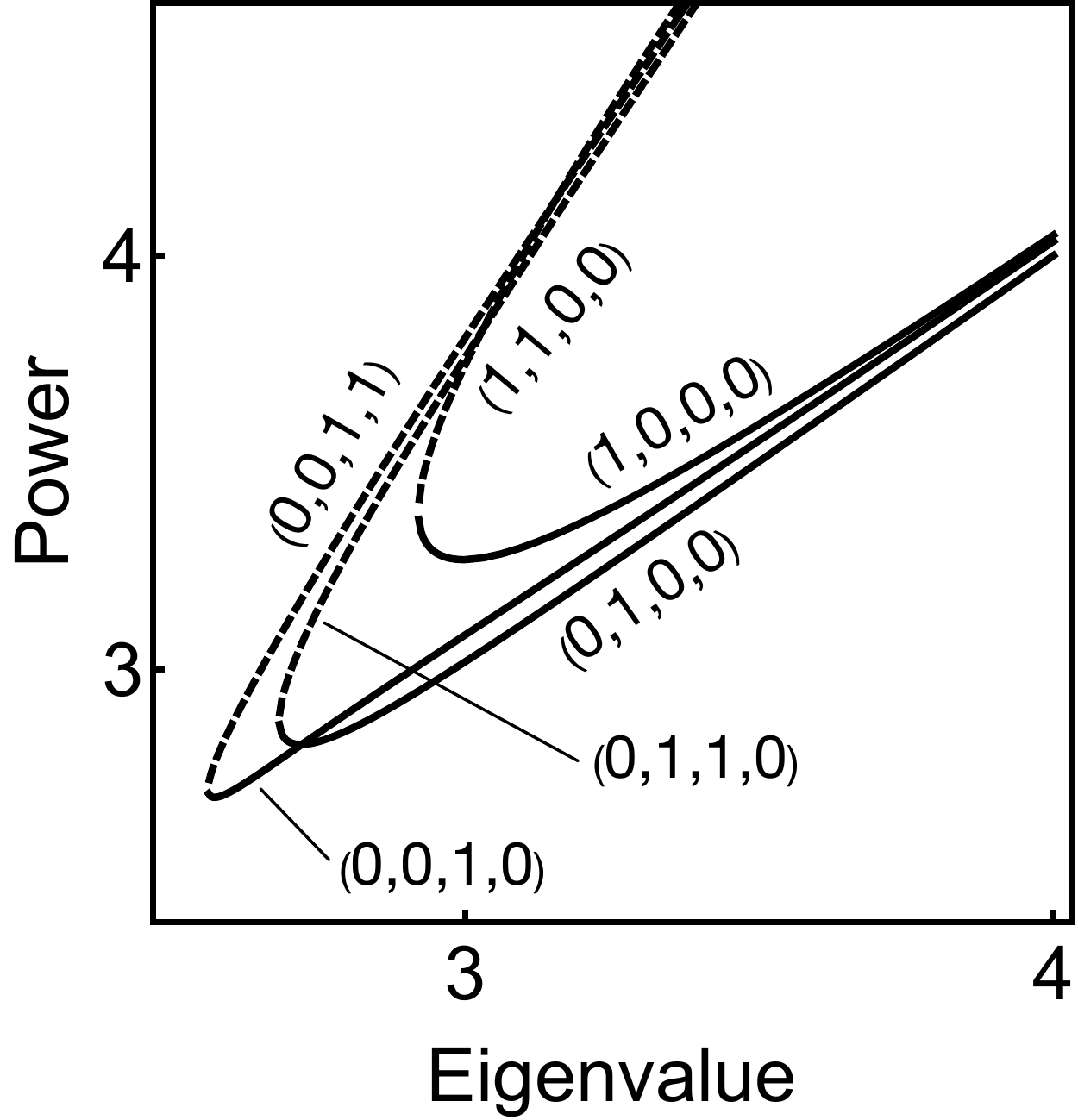}
  \caption{Power content versus eigenvalue for some bulk and surface modes. Continuous (dashed) curves denote stable (unstable) modes.  
 Top left: bulk, $s=1/2$ Top right: bulk, $s=1^{-}$. Bottom left: surface, $s=1/2$. Bottom right: surface, $s=1^{-}$. The labels on each mode denote the anti-continuous state they originated from. Vertical axis scales are different for ease in visualization. ($V = 1, N=51$).
  }
  \label{fig5}
\end{figure}
Figure \ref{fig5} shows the power versus eigenvalues bifurcation curves for $s=1/2$ and $s=1$ (for comparison), for some bulk and surface modes. We note that the states with $s=1/2$ and their counterparts with $s=1$ exhibit general similarities. For instance in both cases the odd and even bulk modes reach all the way down to 
\begin{figure}[b]
 \includegraphics[scale=0.215]{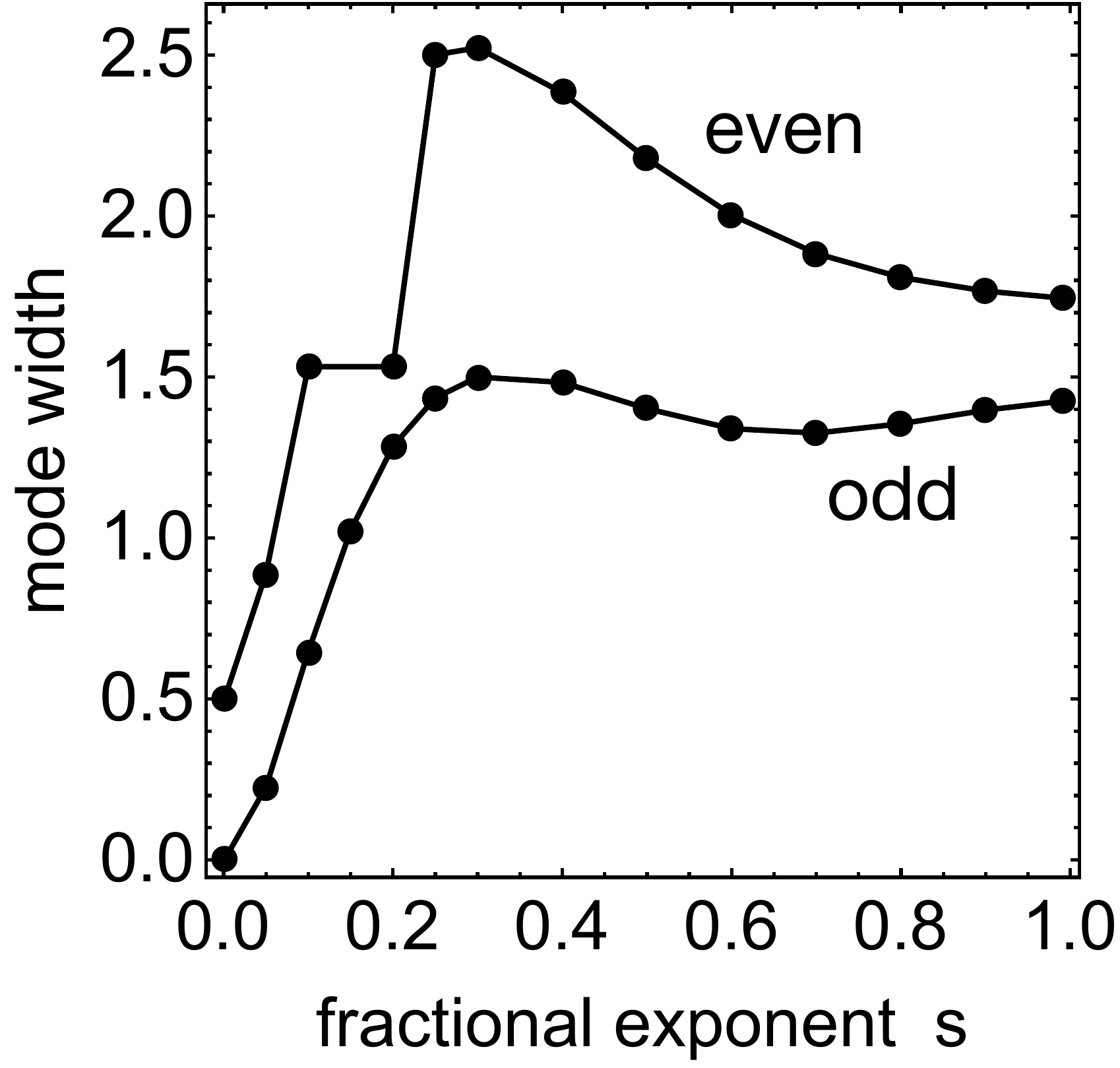}
 \includegraphics[scale=0.255]{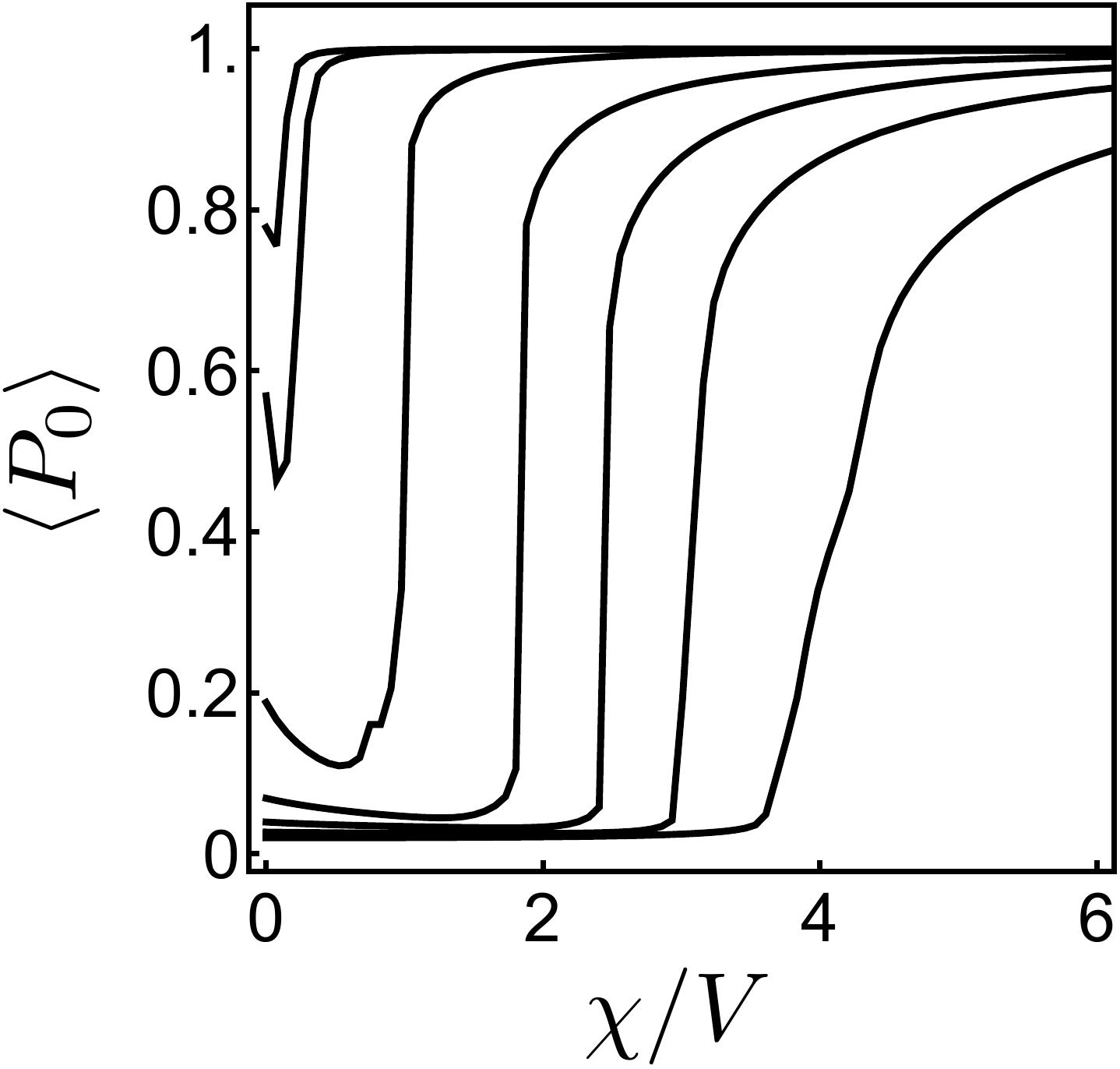}
  \caption{Left: mode width for the nonlinear odd and even modes as a function of the fractional exponent. Right: Time-averaged probability at the initial site as a function of nonlinearity, for several different values of the fractional exponent, $s$. From the leftmost to the rightmost curve $s=0.01, 0.02, 0.1, 0.3, 0.5, 0.7, 0.99$ ($V=1, VT=60$).}
  \label{fig6}
\end{figure}
the band, while for the surface modes, they all need a minimum value of nonlinearity (power) to exist.

{\em Selftrapping}.\ One of the well-known facts about the dnls equation is that it leads to selftrapping, where an initially localized excitation, does not diffuse away completely and a finite fraction remains localized at the initial site. This occurs for a nonlinearity strength greater than a critical value. To find the selftrapping transition, one monitors the time-averaged probability at the initial site,
\be
\langle P_{0} \rangle = {1\over{T}} \int_{0}^{T} |C_{0}(t)|^2 dt
\ee
where $T$ is large and where, without loss of generality the excitation has been initially placed at site $n=0$. We have computed $\langle P_{0} \rangle$ for several $s$ values, comparing the selftrapping curves obtained. Results are shown in Fig.\ref{fig6}, where we see that the existence of a selftrapping transition is preserved for all $s$ values. An interesting feature of these selftrapping curves is that the selftrapping transition moves to smaller values as $s$ is decreased. This can be understood as the effect of the shrinking of the modes as $s$ is decreased (Fig.\ref{fig6}). This mode shrinking facilitates the selftrapping of the excitation, thus decreasing the threshold for trapping. The second interesting feature is the existence of a fraction of linear trapping at $\chi=0$. The amount of trapping increases with a decrease in $s$. As $s$ decreases, this linear trapping approaches unity. A simple way to model this behavior is to assume all sites coupled to each other with identical couplings. The resulting system is known as a simplex\cite{simplex1,simplex2} . It is well-known that when placing an excitation on a given site of the simplex, its time-averaged value is given by
\be
\langle |C_{0}(t)|^2 \rangle = {(N-1)^2+1\over{N^2}}.
\ee
where $N$ is the number of sites. Thus, at large $N$ the trapping fraction approaches unity, as in our case.

\section{Conclusions}
We have examined the effect of replacing the discrete laplacian operator, by a fractional discrete laplacian operator in the dnls equation. In the linear case we have compared the spectra of linear waves, the range of intersite coupling, and the RMS displacement of an initially localized excitation. We found that the bandwidth increases with the fractional exponent $s$, while the RMS is ballistic for all $s$ values, with a non-monotonic propagation `speed'. In the nonlinear case, we examined the bulk and surface modes and their stabilities, for different fractional exponents,  not finding anything substantially different from the dnls case, save for some shifting of the power curves in power-eigenvalue space. The most important influence of the fractional laplacian happens to be  on the dynamics of an initially-localized excitation, where the selftrapping curves shift substantially as a function of $s$, decreasing the selftrapping threshold as $s$ is decreased. In particular, at small $s$ values, linear selftrapping becomes apparent. This fraction of linear selftrapping reaches values close to unity for very small $s$ values. We have attributed this behavior to the extreme long-range of the coupling that exists at very small $s$. From the shape of the dispersion $\lambda(k)$, we see that, at $k=0$, $\lambda=2 V$, independent of $s$. However, for $k\neq0$ $\lambda(k)$ becomes arbitrarily close to $V$, forming a quasi-flat band, with nearly degenerate states. Under those conditions, and based on the simplex analogy, linear trapping is expected on general grounds. The shifting of the selftrapping curves can be understood with the help of Fig\ref{fig6}: As $s$ decreases, the width of the pulse decreases which facilitates trapping, hence a smaller critical nonlinearity is needed.

\acknowledgments
This work was supported by Fondecyt Grant 1160177.

\end{document}